# On materials destruction criteria.

L.S.Kremnev.

Moscow State Technological University "STANKIN", Moscow, Russia,

*Kremnevls@yandex.ru.*

Abstract. In terms of nonlinear material fracture mechanics, the real (discrete)-structure material fracture model has been developed. The model rests on the demonstration of the fact that crack resistance $K_{1c}=2\sigma\sqrt{l}$ and fracture toughness are $G_{1c}=J_{1c}=2\sigma l$ obtained on the basis of energy conservation law and derived without linear material fracture mechanics assumptions can be respectively taken as force and energy criteria for non-linear fracture mechanics. It is shown that $G_{1c}=(K_{1c})^2/E$ is the energy criterion of linear fracture mechanics of material and it is sufficiently less than $G_{1c} = J_{1c}= 2\sigma l$.

It is considered that Elastic Fracture Mechanics (EFA) provides the basis for modern strength analysis of bearing structures and parts. Equation (I) for force criterion ($K_{1c}$) and equation (II) for energy criterion ($G_{1c}$) of material fracture are referred to as the fundamental equations of EFA, these criteria being equal to:

$$K_{1c} = c\sigma\sqrt{\pi\, l}, \qquad (1)$$

$$G_{1c} = K_{1c}^{\,2} / E, \qquad (2)$$

where $K_{1c}$ – the Critical Stress Intensity Factor (CSIF) or crack resistance (CR), $\sigma$ - ultimate stress limit of a specimen having a crack with length $l$, $c$ – factor determined by the specimen form and test conditions, $G_{1c}$ –crack toughness, $E$ – normal elasticity module.

The abovementioned fracture criteria were obtained with the following assumptions taken into consideration:

1. The material being fractured is elastic, i.e. its relative strain is directly proportional to the stress applied. That's why the fracture mechanics based on this assumption is called Elastic Fracture Mechanics, "EFM".
2. The material structure is continuous rather than discrete i.e. atomic.
3. The crack tip radius $\rho \to 0$, and the stress at its tip $\sigma_y \to \infty$.

The listed assumptions put the physical model of fracturing, on the basis of which its force and energy EFM criteria are obtained, rather far from a real material operation and study conditions and its fracture patterns.

Besides, the dimension of $K_{1c}$ [ $Pam^{1/2}$ ], in contrast to the dimensions of other structural and tool material properties' values, fails to adequately reflect its substantial sense. In this regard it would be pertinent to quote J.F. Nott, a prominent expert in the field of Fracture Mechanics: "The physical sense of parameter $K$ is difficult to comprehend mostly due to its dimension (*stress* multiplied by *length square root*), which is rather hard to physically imagine"[1].

Let's consider a plate with thickness $\delta$ made of a not necessarily elastic material acted on by a load with tensile strength $\sigma$ (Fig.1). Let's assume that a side crack having length $l$ and width opening $a$ appeared in the plate. The direct cause of the crack origin is in the fracture stress appearing in its tips ($\sigma_y - \sigma$), whereas the indirect cause consists in the action of stress $\sigma$; $\sigma_y > \sigma$ due to the crack tip acting as a concentrator.

In compliance with the law of energy conservation the values of fracture energies $W_y$ and $W$ of forces $P_y$ and $P$, which account for the appearance of stresses $\sigma_y$ and $\sigma$, are equal for the equilibrium system condition (Fig.1, 2). Let's determine the values of these energies and set them equal.

Energy $W_y$ is determined by:

$$W_y = P_y \cdot S_y, \qquad (3)$$

where $S_y$ –transfer (broadening) of a crack in the direction of $P_y$ along the axis $Y$. From Fig.2 it follows that:

$$P_y = \tfrac{1}{2}(\sigma_y - \sigma) \cdot \delta \cdot \Delta, \qquad (4)$$

where $(\sigma_y - \sigma)$ – fracture stress. Indeed, at $\sigma \to \sigma_y$ it tends to 0, and the plate remains undisturbed, though acted upon by stress $\sigma$. Due to stress $\sigma$ growth and crack development the fracture stress also increases from 0 to $(\sigma_y - \sigma)$. Hence, the average value of the plate fracturing stress is equal to $0.5(\sigma_y - \sigma)$. This circumstance is taken into consideration in (4) by means of factor ½. $\Delta$ - the minimal direction at which the crack edges irreversibly diverge at its tip (Fig.2). Naturally, in the real (discrete) structure material $\Delta$ equals to interatomic spacing (b).

Thus,

$$W_y = ½ (\sigma_y - \sigma) \cdot \delta \cdot \Delta \cdot a, \qquad (5)$$

where $a = \Sigma\Delta = S_y$ (critical crack opening «CCO» at the moment of plate fracturing).

Energy $W$ is determined as

$$W = P \cdot \Delta = \sigma \cdot l \cdot \delta \cdot \Delta. \qquad (6)$$

Having set (5) and (6) equal, we obtain:

$$(\sigma_y - \sigma) \cdot a = 2\sigma \cdot l. \qquad (7)$$

The left-hand part of (7) is known as J. Rice integral $J_{1c}$ [2, 3]. According to J. Rice, the integral for a lateral crack $J_{1c} = \sigma_0 \times \delta_\kappa$, where $\sigma_0$ is a fracture stress at the crack tip, $\sigma_0 = (\sigma_y - \sigma)$ and $\delta_\kappa$ - lateral crack opening ($\delta_\kappa = a$). The right-hand part of (7) constitutes the second form of $J_{1c}$ - integral:

$$J_{1c} = 2\sigma \cdot l \qquad (8)$$

$J_{1c}$ – integral [J/m$^2$] is known to be equal to energy needed to form a unit area of the surface appearing as a result of the crack propagation during elastic - plastic, i.e. real-structure material fracture. Therefore $J_{1c}$ – integral is equal to the fracture toughness $G_{1c}$ [J/m$^2$], i.e. the specific fracture energy or, in other words, the energy criterion of non-linear material fracture mechanics (NLFM). This conclusion also follows from the fact that the forms of $J_{1c}$ - integral (7) and (8) were obtained by the authors on the basis of energy conservation law and without linear fracture mechanics utilization.

Let's consider the fact that from (7) it follows that

$$\sigma_y = \sigma \cdot (1 + 2l/a) \quad (9)$$

and

$$\sigma_y = \sigma \cdot (1 + 2\sqrt{l/\rho}) \quad (10)$$

This equation, as is known from [3], holds for any form of a crack at the contour of which there exists a point with a small curvature radius $\rho$. Equations (9) и (10) are long and well known in the mechanics of real material' fracture and work fairly well.

Let's point out that the value of J-integral $J_{1c} = G_{1c} = 2\sigma \cdot l$ can be calculated by the results of the same standard experiment aimed at determining the force criterion $K_{1c} = 2\sigma \cdot \sqrt{l}$ whereas the value $J_{1c} = G_{1c} = (\sigma_y - \sigma) \cdot a$ is difficult, if ever possible, to be experimentally obtained [4].

Let's put the equation (8) in the form of:

$$G_{1c} = J_{1c} = 2\sigma \cdot l = 2\sigma \cdot \sqrt{l} \sqrt{l} = K_{1c} \cdot \sqrt{l} \quad (11)$$

From equation (11) it follows:

$$K_{1c} = G_{1c} / \sqrt{l} \quad (12)$$

As the cofactor $G_{1c}$ of the right-hand part (12) is the energy criterion of non-linear fracture mechanics, then the force criterion $K_{1c} = 2\sigma \cdot \sqrt{l}$ of the left-hand part is also the criterion of non-linear material fracture mechanics.

Let's remember that the force criterion $K_{1c} = 2\sigma \cdot \sqrt{l}$ was obtained for the case of brittle fracture, i.e. for the conditions of linear fracture mechanics. Nevertheless it is equally successful when used for the case of elastic-plastic fracture, i.e. linear mechanics conditions. This unexpected result is accounted for by the appearance of a thin layer of $\delta_\kappa$- thick plastically deformed material at the crack tip. In the presence of this layer the material transits into the quasi-brittle condition and therefore the asymptotic equations of LFM still hold actual. As a rule the thin layer dimensions $\delta_\kappa$ are limited and never exceed 20% of the crack length [5]. However the value of $K_{1c}$ (Equation (1) ) is widely and successfully used for determining the critical crack length $l_{\text{кр.}}$ of numerous real-structure materials where the crack propagation stage is preceded by forming a sizable zone of plastically deformed material. For example, high-tempered structural steels fall into this category of materials.

Linear dimension ▼ of the plastic deformation zone at the edge crack tip, which is observed immediately before its propagation under the plane-strain condition, is equal to: [4]:

$$▼ = (1/3\pi)(K_{1c}/\sigma_\text{т})^2. \tag{13}$$

From $K_{1c} = 2\sigma\sqrt{l}$ we obtain:

$$l_\text{кр.} = 0{,}25 \cdot (K_{1c}/\sigma_\text{в})^2. \tag{14}$$

Using (13) and (14) we can find the relative dimension of ▼ plastic deformation zone at the side crack tip immediately before the material fracture:

$$▼ / l_\text{кр} = 0{,}42\,(\sigma_\text{в}/\sigma_T)^2. \tag{15}$$

By means of equation (15) we can assess ▼/$l_\text{кр}$ of a common structural steel 40X. After standard heat treatment the hardness of Steel 30 – 32HRC, $\sigma_\text{B}$ and $\sigma_{0.2}$ are equal to 1030MPa and 800MPa [6] respectively. In compliance with equation (15) the extension of the plastic zone at the crack type in steel 40X accounts for 70% of its length. In case of Steel 3 the dimension of this experimentally revealed zone [6] notably exceeds the crack length. The provided values of ▼/$l_\text{кр}$ considerably exceed the acceptable value of the thin layer $\delta_к$, which as has been found fails of exceed 20%$l_\text{кр}$.

Therefore the explanation to the fact that the laws of LFM hold for elastic-plastic materials due to their transition to the quasi-brittle condition taking place as a consequence of a thin deformed layer $\delta_к$ appearing at its tip looks insufficiently convincing. In contrast to the abovementioned the obtained proof explaining the fact that the force fracture criterion $K_{1c}$(CR) (I) represents the criterion of both elastic and elastic-plastic material fracture, i.e. NLFM, explicitly clarifies the situation.

Bearing in mind that $K_{1c} = 2\sigma \cdot \sqrt{l}$, we obtain from (11):

$$G_{1c} = J_{1c} = 2\sigma \cdot l = (K_{1c})^2 / 2\sigma. \tag{16}$$

The obtained energy criterion of non-linear material fracture mechanics is by several orders of magnitude larger than the energy criterion of linear fracture mechanics $G_{1c} = (K_{1c})^2/E$ (2). Foe example, for Steel 40X after hardening with subse-quent drawing $\sigma_B \sim 1000$ MPa, and normal module of elasticity $E \sim 220\,000$ MPa. Hence, $G_{1c}$ ($J_{1c}$) is approximately by 100 times larger than $G_{1c} = (K_{1c})^2/E$. At the same time the fracture process consideration by means of the crack model presented (Fig.1, 2) provides the possibility to exclude the assumptions of the fact that the radius of the crack tip curvature in the plate made of a real-structure material $\rho \to 0$, and the stress at its tip $\sigma_y \to \infty$. Sure enough, in such a material $\rho$ cannot be lower than the interatomic space $b$, and $\sigma_y$ -- higher than the ultimate strength limit $\sigma_{teor} \sim 0,1E$. As a consequence of the abovementioned it turned out that the crack-resistance of a material with obtuse cracks is by 13% higher than that in the case of an acute crack: $1,13\sigma\sqrt{\pi l}$ [equation (13)] and $\sigma\sqrt{\pi l}$ [7] respectively]. These should be considered, if the values of $K_{1c}$ presented in reference sources, refer to the cracks with $\rho \to 0$.

From equation (12) one can find the physical dimension for the force criterion of NLFM: $K_{1c}$:[J/m$^2$]/m$^{1/2}$. Thus, the physical sense of the force NLFM criterion crack resistance $K_{1c}$ is not obvious. Along with these the author thinks it expedient to generally turn from the force criterion to a more easily comprehended energy criterion NLFM crack toughness $G_{1c} = J_{1c} = 2\sigma l$. Let us recall that the results of a single standard experiment make it possible to determine both force and energy NLFM criterion.

Conclusions:

1. A model of crack propagation referring to the real (discrete) - structure material has been offered.
2. The model is based on Non-Linear Fracture Mechanics (NLFM) equation establishing the force and energy criteria for real-structure materials' fracturing.
3. It has been shown that the crack resistance $K_{1c} = 2\sigma \cdot \sqrt{l}$ itself as well as the fracture toughness $G_{1c} = J_{1c} = 2\sigma \cdot l = (K_{1c})^2 /2\sigma$ can be respectively considered as force and energy criteria of non-linear material fracture mechanics.

4. The criterion $G_{1c} = (K_{1c})^2/E$ constitutes the energy criterion on Linear Fracture Mechanics (LFM) and its value is considerably lower than that of $G_{1c} = J_{1c}$.

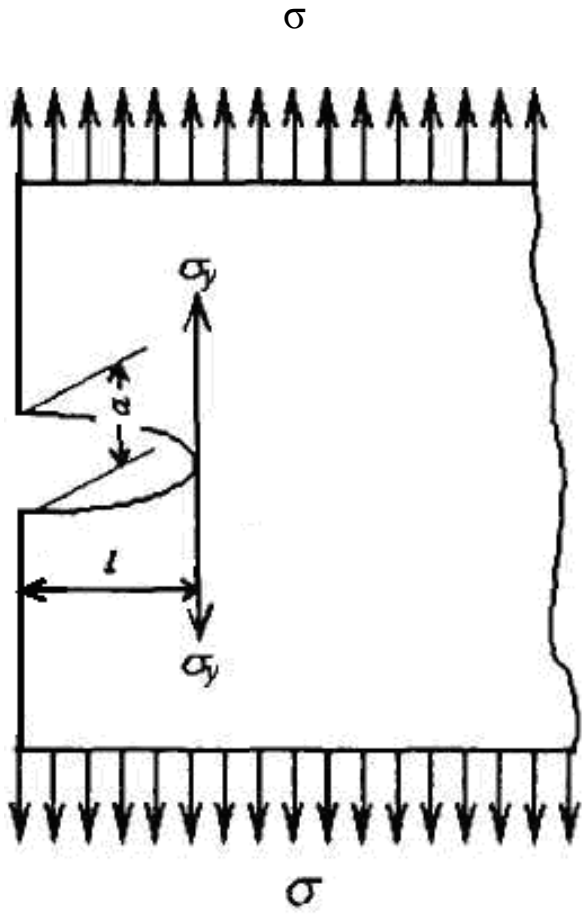

**Fig.1**

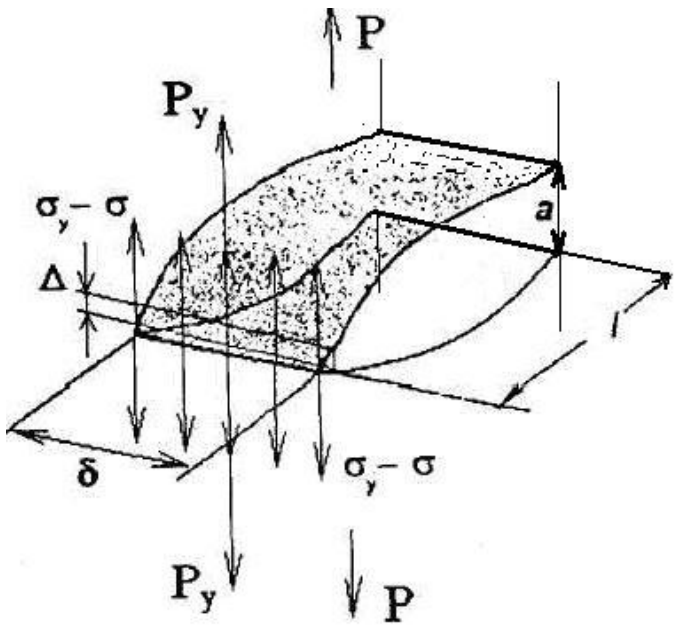

**Fig.2**

Figure captions:

Fig.1. A $\delta$- thick plate with an $l$- long and $a$ - width lateral crack acted by loading with stress σ; $(\sigma_y - \sigma)$ is the crack tip stress; $\sigma_y > \sigma$.

Fig.2. Fig.1 crack pattern; Δ - minimal irreversible crack divergence at its tip taking place under the action of ultimate stress $(\sigma_y - \sigma)$.